\documentclass{aa}
\usepackage{graphicx}
\usepackage{natbib}
\usepackage{aas_macros}
\bibpunct{(}{)}{;}{a}{}{,}

\newcommand{\msun}{\mbox{${\rm M}_{\sun}$}}

\newcommand{\rsun}{\mbox{${\rm R}_{\sun}$}}

\begin{document}

\title{The gravitational wave signal from the Galactic disk population of
  binaries containing two compact objects}

\author{G. Nelemans\inst{1}, L. R. Yungelson\inst{1,2} and S. F.
  Portegies Zwart\inst{3}\thanks{Hubble Fellow}} \offprints{G. Nelemans, GijsN@astro.uva.nl}

\institute{Astronomical Institute ``Anton Pannekoek'', 
        Kruislaan 403, NL-1098 SJ Amsterdam, the Netherlands 
             \and
             Institute of Astronomy of the Russian Academy of
             Sciences, 48 Pyatnitskaya Str., 109017 Moscow, Russia 
                \and 
        Massachusetts Institute of Technology, Massachusetts Ave. 77,
                Cambridge MA 02139, USA
}

\date{accepted 2001 May 7}

\titlerunning{Gravitational waves from binaries with two compact objects}
\authorrunning{Nelemans, Yungelson \& Portegies Zwart}

\abstract{We review the properties of Galactic binaries containing two
  compact objects, as derived by means of population synthesis.  Using
  this information we calculate the gravitational wave signal of these
  binaries. At frequencies below $f \la 2$~mHz the double white dwarf
  population forms an unresolved background for the low-frequency
  gravitational wave detector LISA. Above this limit some few thousand
  double white dwarfs and few tens of binaries containing neutron
  stars will be resolved. Of the resolved double white dwarfs
  $\sim500$ have a total mass above the Chandrasekhar limit. About
  $\sim95$ of these have a measurable frequency change allowing a
  determination of their chirp mass. We discuss the properties of the
  resolved systems.
\keywords{Gravitational waves -- Stars: statistics --  binaries: close -- Galaxy: stellar content }
}

\maketitle

\section{Introduction}

The interest in gravitational waves, predicted by Einstein's theory of
general relativity, was greatly enhanced by the signals supposedly
detected by resonant gravitational wave (GW) antennas \citep{web69}
and the discovery of the pulsar B1916+13 in a relativistic binary
\citep{ht75,tw82}. Currently, about ten projects for ground and
space-based gravitational wave detectors are already operating or
under development \citep[see][]{fla98}. They will open the windows in
the frequency bands 10 to $10^4$\,Hz from the ground and $10^{-4}$ to
1\,Hz from space. Recently the first upper limits on detections from
the Japanese TAMA300 detector were reported \citep{tkt+00}.

At high frequencies the merging events of extragalactic binaries
containing neutron stars and/or black holes are among the most
promising sources of GW radiation.  The estimates of the merger rates
of these systems are highly uncertain \citep[e.g.][]{phi91,py98,kl00}.
An upper limit for the rate of neutron star -- neutron star mergers in
our Galaxy of $\sim10^{-4}\,{\rm yr^{-1}}$ is found both from observations
\citep{acw99} and theory \citep{ty93b}. Extrapolated to cosmic scales,
these estimates show that the perspectives for detection of such
events by the first generation GW detectors are not very good
\citep[see][]{kns+00}.  They could be better for black hole -- black
hole or for black hole -- neutron star mergers
\citep{ty93b,lpp97,pm00}.

At low frequencies, it was first expected that contact W~UMa binaries
will dominate the gravitational wave spectrum \citep{mir65}. However,
it was shown that the gravitational wave background formed by Galactic
disk systems is probably totally dominated by detached double white dwarfs and
that their number is so large that they will form a confusion limited
background for the currently planned detectors
\citep{eis87,lpp87,hbw90,npv00}. Only sources with a frequency above a
certain limiting frequency (somewhere between $\sim\!1 - 10$mHz) can
be resolved \citep{eis87}.
 
The aim of the present paper is an accurate evaluation of the
confusion limit, based on population synthesis models for compact
stars in the Galactic disk and a discussion of the properties of the
sample of potentially resolved binaries containing two compact
objects: white dwarfs, neutron stars or black holes. 
%The Galactic disk populations of close binaries with compact objects
%were modelled in our previous studies \citep{py98,py99,nyp+00,npv+00}.
We first discuss the gravitational wave signal from (eccentric)
binaries (Sect.~\ref{GWR}). Next, we summarise the properties of the
Galactic disk populations of compact binaries which are relevant to the
emission of gravitational waves (Sect.~\ref{population}). We do not
consider globular cluster binaries. In Sect.~\ref{results} we present
a model for the background formed by the Galactic disk double white
dwarfs, discuss the confusion limit and the properties of the
individually resolved binaries. A discussion of the possible
contribution of the halo and extragalactic sources and a comparison
with previous work follows in Sect.~\ref{discussion}. Our conclusions
are summarised in Sect.~\ref{conclusion}.

\section{Gravitational waves from binaries}\label{GWR}

The gravitational wave luminosity of a binary in the $n$th harmonic is
given by \citep{pm63} 
\begin{equation}\label{eq:L}
L (n, e) = \frac{32}{5} \, \frac{G^4}{c^5} \frac{\,M^2 \,m^2
  \,(M+m)}{a^{5}} g(n, e).
\end{equation}
Here $M$ and $m$ are the masses of the components, $a$ is their
orbital separation and $e$ is the eccentricity of the orbit. The
function $g(n, e)$ is the Fourier decomposition of the GW signal.

\begin{figure}[t]
\resizebox{\columnwidth}{!}{\includegraphics{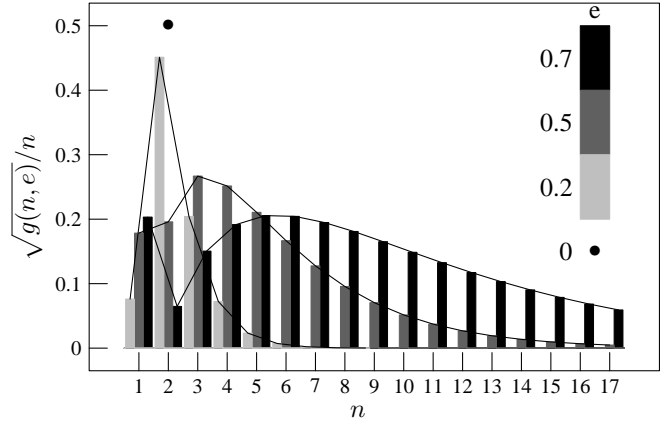}}
\caption[]{Scale factor of the GW strain amplitude $\sqrt{g(n,
    e)}/n$ for the different harmonics [Eq.~(\ref{eq:h})] for e = 0,
  0.2, 0.5 and 0.7.  }
\label{fig:g}
\end{figure}

The measurable signal for gravitational wave detectors is the
amplitude of the wave -- $h_+$ and $h_{\times}$ for the two
polarisations. These can be computed from the GW flux at the Earth
\citep{pt72}
\begin{equation}
 \frac{L_{\rm GW}}{4 \pi d^2} = F = \frac{c^3}{16 \pi G} \langle \dot{h}_+^2 + \dot{h}_{\times}^2 \rangle.
\end{equation}
Assuming the waves to be sinusoidal and defining the so called strain
amplitude as $h = ({1 \over 2} [h_{\rm +,max}^2 +
h_{\rm \times,max}^2])^{1/2}$ one obtains
\begin{eqnarray}\label{eq:h}
& &h (n, e) = \left[ \frac{16 \pi \, G}{c^3 \, \omega_{\rm g}^2} 
   \frac{L(n, e)}{4 \pi \, d^2} \right]^{1/2}  \\ 
& & = 1.0 \,10^{-21}  \frac{\sqrt{g(n,e)}}{n}  \left( \frac{\mathcal{M}}{\msun}
\right)^{5/3} \! \! \! \left( \frac{P_{\rm orb}}{\rm 1\, {\rm hr}} \right)^{-2/3}
\! \! \!\left( \frac{d}{\rm 1 {\rm kpc}} \right)^{-1}, \nonumber
\end{eqnarray}
where $\mathcal{M} = (M \, m)^{3/5} / (M + m)^{1/5}$ is the so called
chirp mass and $\omega_{\rm g} = \pi n/P_{\rm orb}$ is the angular
frequency of the emitted wave\footnote{For circular orbits this
  equation is identical to Eq.~(5) of \citet{eis87}. It is different
  by a factor of $\sqrt{8}$ from Eq.~(20) of \citet{pt72}, who use a
  factor $\sqrt{2}$ larger definition of $h$ and possibly confuse
  $\omega_{\rm g}$ in Eq.~(\ref{eq:h}) with the orbital angular
  frequency.  It differs by a factor $2^{5/3}$ from Eq.~(3.14) of
  \citet{db79} because they confuse the orbital frequency in their
  Eq.~(3.13), with the frequency of the wave (twice the orbital
  frequency) in their Eq.~(3.4).}.  In Fig.~\ref{fig:g} we plot the
values of $\sqrt{g(n, e)}/n$ for different eccentricities. High
eccentricity binaries emit most of their energy at higher frequencies
than their orbital frequency, reflecting the fact that the radiation
is more effective near periastron of the orbit. Thus, eccentric
compact binaries may be detectable sources of GW signals at
frequencies higher than their orbital frequency
\citep[cf.][]{bmp+88,hil91}.

\section{The Galactic disk population of binaries containing two compact
  objects}\label{population}

We calculated the Galactic disk population of binaries containing two
compact objects using the population synthesis code \textsf{SeBa}
\citep{pv96,py98,nyp+00}. The basic assumptions used in this paper can
be summarised as follows. The initial primary masses are distributed
according to a power law IMF with index $-2.5$, the initial mass ratio
distribution is taken flat, the initial semi major axis distribution
flat in log $a$ up to $a = 10^6 \rsun$, and the eccentricities follow
$P(e) \propto 2 e$. The fraction of binaries in the initial population
of main-sequence stars is 50\% (2/3 of all stars are in binaries). A
difference with other studies of the populations of close binaries is
that the mass transfer from a giant to a main sequence star of
comparable mass is calculated using an angular momentum balance
formalism, as described in \citet{nvy+00}.  For the
star formation rate of the Galactic disc we use an exponential
function:
\begin{equation}\label{eq:sfr}
{\rm SFR}(t) = 15 \; \exp(-t/\tau) \quad \msun \; \mbox{yr}^{-1},
\end{equation}
where $\tau$ = 7 Gyr. With an age of the Galactic disk of 10\,Gyr it
gives a current star formation rate of 3.6\,\msun\,yr$^{-1}$
compatible with observational estimates \citep{ran91,hj97}. It gives a
Galactic supernova II/Ib rate of 0.02 yr$^{-1}$ and if supernovae Ia
are produced by merging double carbon-oxygen (CO) white dwarfs it
gives a Galactic rate of 0.002 yr$^{-1}$. Both are in agreement with
observational estimates by \citet{cet99}.

\begin{table}[t]
\caption[]{Current 
  birth rates ($\nu$) and merger rates ($\nu_{\rm merg}$) per year for
  Galactic disk binaries
  containing two compact objects and their total number ($\#$) in the
  Galactic disk, as calculated with the \textsf{SeBa} population
  synthesis code (see text).  
}
\label{birthrates}
\begin{tabular}{lccc}
 Type  &  \multicolumn{1}{c}{$\nu$}  & \multicolumn{1}{c}{$\nu_{\rm merg}$}  & \multicolumn{1}{c}{$\#$}     \\ \hline      
\raisebox{0pt}[2.5ex]{}(wd, wd)  &  2.5 $\! \times \!$ 10$^{-2}$ & 1.1 $\! \times \!$ 10$^{-2}$ & 1.1 $\! \times \!$ 10$^{8}$ \\
 $[$wd, wd)&  3.3 $\! \times \!$ 10$^{-3}$ & --                            & 4.2 $\! \times \!$ 10$^{7}$ \\
 (ns, wd)  &  2.4 $\! \times \!$ 10$^{-4}$ & 1.4 $\! \times \!$ 10$^{-4}$ & 2.2 $\! \times \!$ 10$^{6}$ \\
 (ns, ns)  &  5.7 $\! \times \!$ 10$^{-5}$ & 2.4 $\! \times \!$ 10$^{-5}$ & 7.5 $\! \times \!$ 10$^{5}$ \\
 (bh, wd)  &  8.2 $\! \times \!$ 10$^{-5}$ & 1.9 $\! \times \!$ 10$^{-6}$ & 1.4 $\! \times \!$ 10$^{6}$ \\
 (bh, ns)  &  2.6 $\! \times \!$ 10$^{-5}$ & 2.9 $\! \times \!$ 10$^{-6}$ & 4.7 $\! \times \!$ 10$^{5}$ \\
 (bh, bh)  &  1.6 $\! \times \!$ 10$^{-4}$ & --                            & 2.8 $\! \times \!$ 10$^{6}$ \\
 \hline 
\end{tabular}
\end{table}

The current birth- and merger rates and total number of systems in the
Galactic disk with these assumptions are given in
Table~\ref{birthrates}. We use a notation introduced by \citet{pv96}:
wd, ns and bh for white dwarf, neutron star and black hole
respectively; $( \quad)$ and $[ \quad )$ for detached and
semi-detached binaries. The fact that the numbers here are different
from the numbers given in \citet{py98,py99}\footnote{Note that
  \citet{py99} consider only a subset of the (ns, wd) population,
  namely the systems that are \textit{eccentric} and contain a white
  dwarf more massive than 1.1\,\msun. For this subset they find 
  a birth rate that is comparable to the birth rate of double neutron
  stars.}, \citet{nyp+00} and \citet{blp+01} is caused by the
differences in the assumed IMF, initial binary fraction and star
formation history.

\begin{figure}[t]
\resizebox{\columnwidth}{!}{\includegraphics[angle=-90,clip]{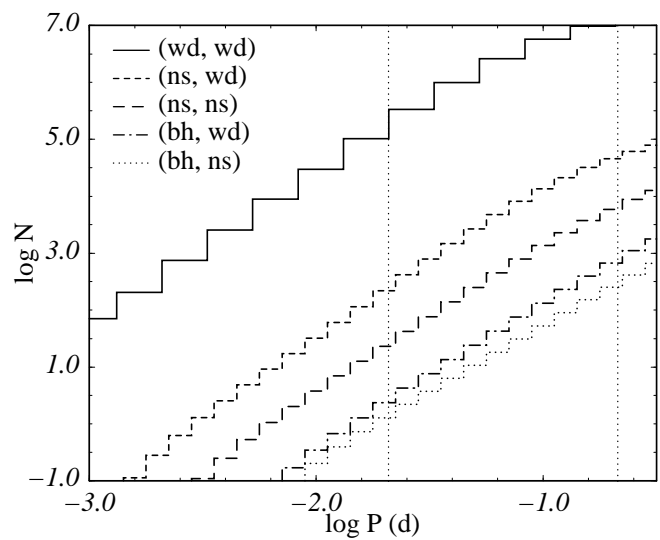}}
\caption[]{Period distribution of the binaries  of different
  types in the period range of interest to the space-based
  gravitational wave detectors like LISA. The vertical dotted lines
  give the periods at which the frequency of the fundamental ($n = 2$)
  harmonic of the gravitational wave is 1 and 0.1 mHz respectively.}
\label{fig:logP_logN}
\end{figure}

In Fig.~\ref{fig:logP_logN} we show the period distributions of the
binaries of different types in the range of interest for space based
gravitational wave detectors. The properties of these populations can
be summarised as follows:\\
\textbf{Detached double white dwarf binaries: (wd, wd)}.
Our model for the Galactic disk population of double white dwarfs is
described in detail in \citet{nyp+00}. Most double white dwarfs have a
mass ratio around unity and low-mass ($M < 0.45 \msun$) components.
From Table~\ref{birthrates} and Fig.~\ref{fig:logP_logN} it is clear
that they vastly outnumber all other binaries with compact objects in
the Galactic disk.\\
\textbf{Semi-detached double white dwarfs (AM CVn stars): [wd, wd)}.
We include in our calculation both AM CVn stars descending from
detached close double white dwarfs and from low-mass helium stars with
white dwarf companions \citep{npv+00}. We use Model II of Nelemans et
al., which is most favourable for the formation of AM CVn's.\\
\textbf{Neutron star -- white dwarf binaries: (ns, wd)}.
Neutron star - white dwarf binaries fall into two families
\citep{ty93c,py99,ts00}. In one family the neutron star is formed
first. Later the secondary forms a white dwarf and in the mass
transfer event the orbit circularizes \citep[e.g.][]{ht84}.  If both
components of the initial binary are of comparable mass it can happen
that the primary becomes a white dwarf, while the secondary accretes
so much mass that it becomes a neutron star \citep[e.g.][]{ty93c}.  In
this case the orbits are eccentric.  The masses of the white dwarfs
are typically low in the first family and high in the second (see
Fig.~\ref{fig:fMch} below). \\
\textbf{Double neutron star binaries: (ns, ns)}.
The formation and characteristics of the current population of double
neutron stars is extensively studied by us in \citet{py98}. Maybe the
most important assumption, which influences the birth rate, orbital
periods and eccentricities of neutron star -- neutron star binaries,
is the kick velocity distribution. We use the one proposed by \citet{har97b}.\\
\textbf{Black hole binaries: (bh, wd), (bh, ns) and (bh,~bh)}.
The knowledge of the way in which black holes are formed and the range
of masses of their progenitors are still highly uncertain
\citep[see e.g.][]{ww95,pve97,eh98a,wl99,fry99}. The treatment of the
formation of black holes implemented in the present study is described
in some detail in the Appendix. Typical black holes in our model have
masses between 5 and 7 \msun. In the short orbital period range
(Fig.~\ref{fig:logP_logN}) they are rare and their merger rates are at
least an order of magnitude lower than those of the neutron star
binaries (Table~\ref{birthrates}). Double black hole binaries are
absent in this period range and do not merge at all in our model.

\begin{figure*}[t]
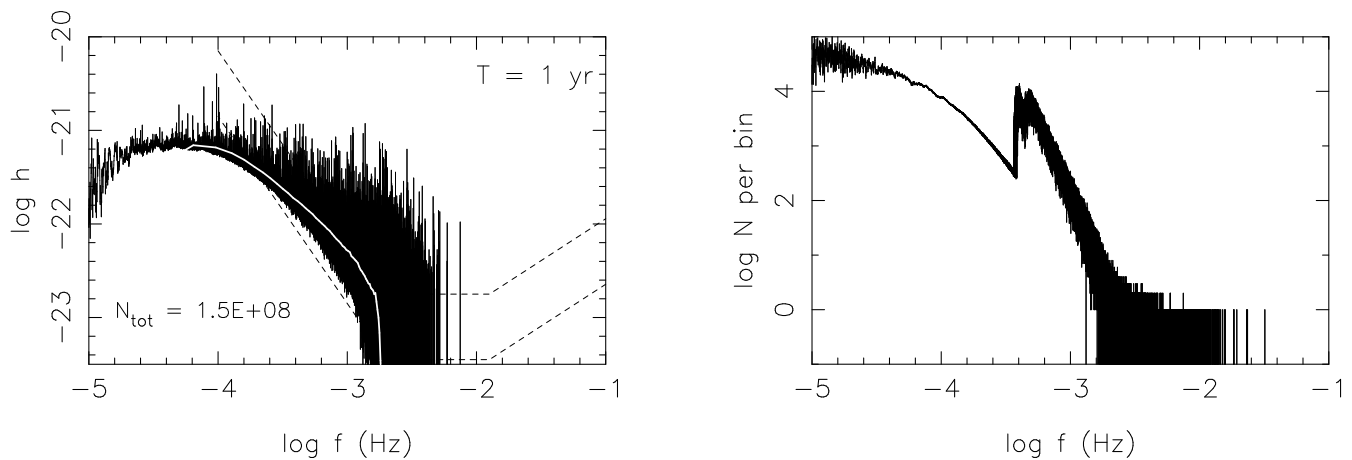

\begin{minipage}{0.53\textwidth}
\resizebox{0.85\textwidth}{!}{\includegraphics[angle=-90]{h2754f3.ps}}
\end{minipage}
\begin{minipage}{0.53\textwidth}
\resizebox{0.85\textwidth}{!}{\includegraphics[angle=-90]{h2754f4.ps}}
\end{minipage}
\caption[]{
  \textbf{Left:} GWR background produced by double white dwarfs (both
  detached and semi-detached). The assumed integration time is 1 yr.
  The `noisy' black line gives the total power spectrum, the white
  line the average. The dashed lines show the expected LISA
  sensitivity for a S/N of 1 and 5. \textbf{Right:} The number of
  systems per bin on a logarithmic scale. The contribution of the
  semi-detached double white dwarfs between $\log f \simeq -3.4$ and $-3.0$
  is clearly visible.  }
\label{fig:GWR_bg}
\end{figure*}

\section{The gravitational wave signal from compact binaries in the Galactic disk }\label{results}

Merging of binaries containing neutron stars and black holes in
distant galaxies could give measurable signals in the high frequency
detectors. We do not extrapolate our merger rates to extragalactic
scales, but our inferred rates (Table~\ref{birthrates}) are consistent
with the (very uncertain) rates derived elsewhere for the Galaxy
\citep[see for a detailed discussion][]{kns+00}.

The Galactic binaries with periods less than 10 hr are interesting for
the low-frequency GW detectors. We calculate the expected signal for
LISA, the joint ESA, NASA detector that is expected to be launched
around 2010. It will consist of three satellites, 5 million kilometres
apart, between which laser beams will be exchanged, measuring the
distance changes \citep{mwh00}. It will give the GW amplitude as
a function of frequency with fixed frequency resolution. A limited
angular resolution will be achieved, allowing e.g. identification of
sources in globular clusters \citep[see][]{bpr01}. In our calculations we
restrict ourselves to the sensitivity in frequency and do not consider
the angular resolution.

Because the number of Galactic binaries drops strongly towards shorter
periods (Fig.~\ref{fig:logP_logN}) the number of sources per frequency
bin for detectors with a fixed frequency resolution will also
decrease: at low frequencies the signals in particular frequency bins
will overlap, forming a so called ``confusion limited noise''. Above a
certain limiting frequency, called the ``confusion limit'', there is
not more than one system per frequency bin, so the systems can be
resolved individually. We discuss these regimes separately.

\subsection{The confusion limited background due to double white dwarfs}

\citet{eis87} have shown that for space-born detectors the confusion
limit is determined by the Galactic close binary white dwarfs. In our
model the total number of detached and semi-detached double white
dwarfs in the Galactic disk is $1.5 \times 10^8$ (see
Table~\ref{birthrates}). We distribute these systems randomly in the
Galactic disk according to
\begin{equation}\label{eq:rho}
        \rho(R, z) = \rho_{\rm 0} \; e^{-R/H} \; 
                     \mbox{sech}(z/\beta)^2  \quad \mbox{pc}^{-3}.
\end{equation}
Here $H$ = 2.5 kpc \citep{sac97} and $\beta$ = 200 pc, neglecting the age
and mass dependence of $\beta$. All systems are circular and for each
system we calculate the strain amplitude from Eq.~(\ref{eq:h}) taking
$R_{\sun} = 8.5$~kpc and $z_{\sun} = -30$~pc.

To simulate the power spectrum for this population of binaries as
would be detected by a gravitational wave detector in space we
determine the distribution of the systems over $\Delta f = 1/T$ wide
bins, with $T$ the total integration time (for which we use 1 yr). In
Fig.~\ref{fig:GWR_bg} we plot the resulting confusion limited
background signal and the number of systems per bin. The contribution
of the semi-detached double white dwarfs, which are less numerous than
the detached double white dwarfs and have lower strain amplitude is
concentrated in a relatively small frequency interval between $\log f
\simeq -3.4$ and $-3.0$ where they dominate the number of systems per
bin.

Most previous studies used only the average properties of the double
white dwarf population to calculate the average background signal. In
Fig.~\ref{fig:GWR_bg} we plotted the average of our model power
spectrum as the white line. Note that in many bins the actual power is
much larger than the average. These bins contain one system that has a
much stronger signal than the rest, for example because it is close to
the Earth, and it may be detectable above the noise (see
Sect.~\ref{detectable}).

\begin{figure}[t]
\resizebox{\columnwidth}{!}{\includegraphics[angle=-90]{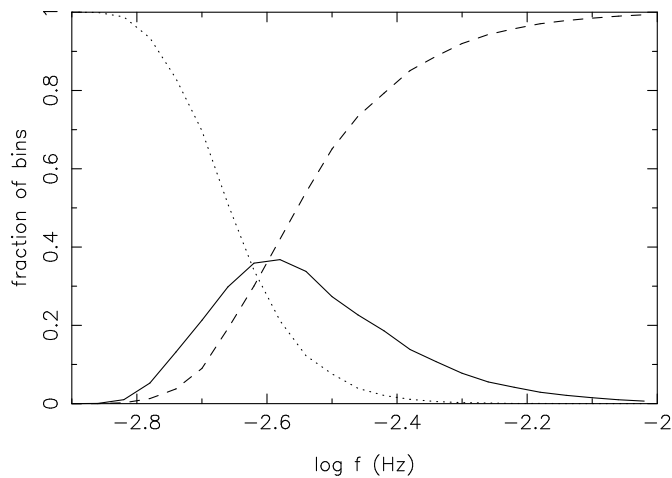}}
\caption[]{Fraction of bins that contain exactly one system
  (solid line), that are empty (dashed line) and contain more than one
  system (dotted line) as function of the frequency of the signals.
  For all frequency intervals the result is normalised, because the
  total number of bins in a logarithmic interval changes strongly.}
\label{fig:binfraction}
\end{figure}

\begin{figure*}[t]
\resizebox{\textwidth}{!}{\includegraphics[angle=-90]{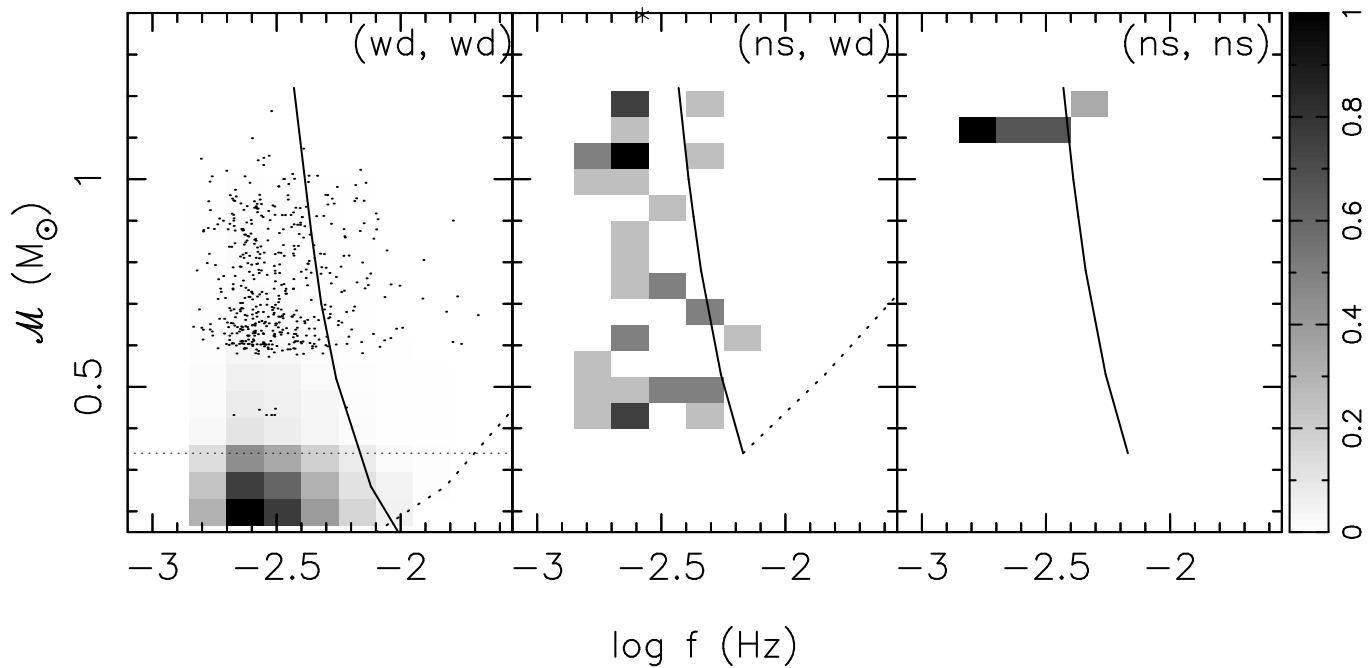}}
\caption[]{
  Distribution of the resolved binaries over frequency and chirp mass
  $\mathcal{M}$. The grey shade gives the number of systems in each
  bin relative to the maximum in each plot which is 1735 for (wd, wd),
  5 for (ns, wd) and 3 for (ns, ns). The type of binary is denoted at
  the top. The asterisk indicates the (bh, wd) system. The dots in the
  (wd, wd) panel indicate the systems with a total mass above the
  Chandrasekhar limit. The solid line shows the chirp line for $T =
  1$~yr (Sect.~\ref{resolved}). The curved dotted line shows the
  position at which the systems merge. Double neutron stars merge at
  high ($f \approx$~kHz) frequencies so their merger line falls off
  this plot.}
\label{fig:fMch}
\end{figure*}

\subsection{The population of resolved binaries}\label{resolved}

Given the fact that the double white dwarf background buries all
underlying signals at frequencies below $\log f \, \approx -2.8$, we
did not consider the neutron star and black hole binaries below 1 mHz.
To find the binaries that will be resolved by LISA, we calculated the
Galactic disk population of all binaries containing compact objects
which contribute to the GW signal at frequencies above 1\,mHz.
Because we now also consider eccentric binaries, which emit at
frequencies higher than twice the orbital frequency
[Eq.~(\ref{eq:h})], there are contributions from binaries with
orbital periods up to $\sim$10\,hr.

If one considers an average background as done i.e. by \citet{eis87},
the average number of systems per bin at a certain moment drops below
one system.  However, we model all individual systems in the Galaxy
and determine, for an integration time $T$ of 1 yr, for each frequency
bin ($\Delta f = 1/T$) how many systems it contains.  In
Fig.~\ref{fig:binfraction} we plot the fraction of bins that contain
exactly one, none and more than one system as function of frequency.
The figure shows that the notion of a ``confusion limit'' as a unique
value is too simple. At $\log f = -2.84$ the first resolved bins (i.e.
containing exactly one system) are found, while up to $\log f = -2.3$
bins containing more than one system are still present.

\begin{table}[t]
\caption[]{The number of resolved systems of different types
  (see Sect.~\ref{resolved}) and the
  number of strong signal systems potentially  detectable above the
  double white dwarf background (see Sect.~\ref{detectable}).
}
\label{tab:resolved}
\begin{tabular}{lrr}
Type             & resolved systems  & detectable above noise   \\ \hline 
 (wd, wd)        & 12124          & 5943       \\
 (ns, wd)        & 38             & 124        \\
 (ns, ns)        & 8              & 31         \\
 (bh, wd)        & 1              & 3         \\
 (bh, ns)        & 0              & 3         \\ \hline
total            & 12171          & 6104     \\ \hline
\end{tabular}
\end{table}

\begin{figure*}[t]
\resizebox{\textwidth}{!}{\includegraphics[angle=-90]{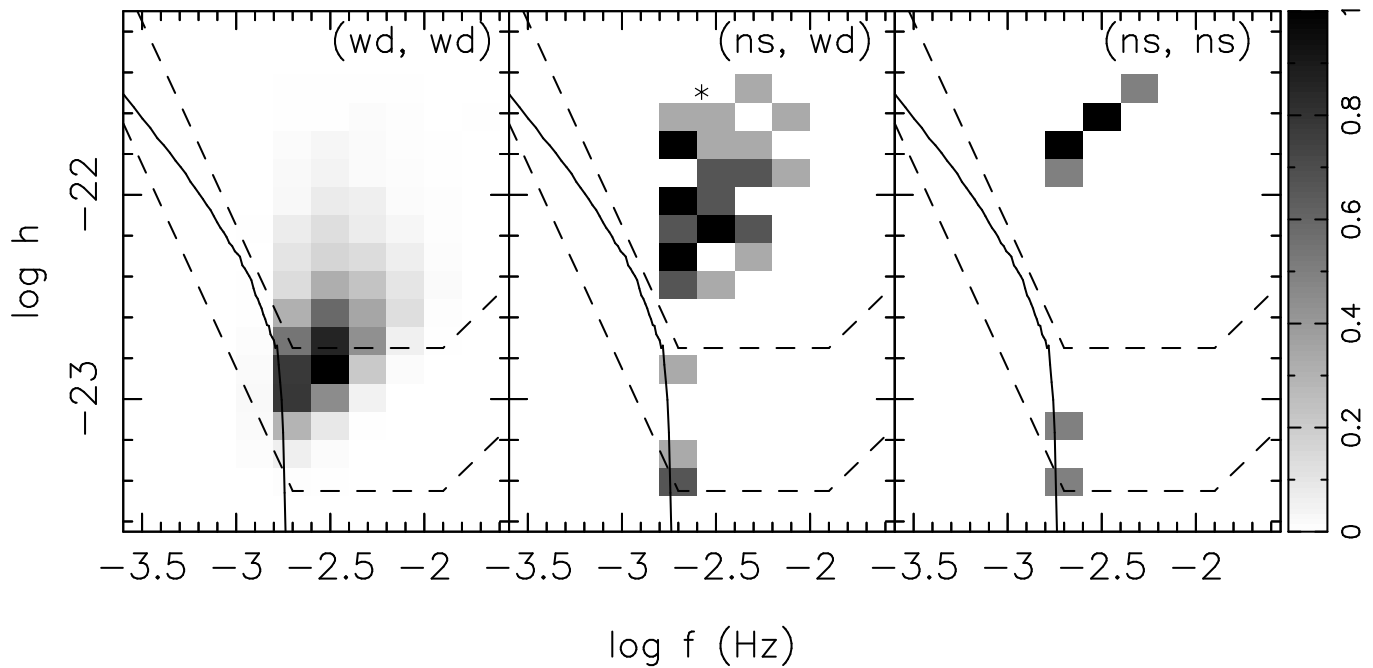}}
\caption[]{
  Distribution of resolved systems over the frequency and strain
  amplitude for the different types of binaries (indicated in the top
  right corner of each panel). The grey shade gives the number of
  systems relative to the maximum in each plot, which is 1314 for (wd,
  wd), 3 for (ns, wd) and 2 for (ns, ns). The asterisk indicates the (bh,
  wd) system.  The dashed lines give the LISA sensitivity for an
  integration time of 1 yr and a signal to noise ratio of 5 (top line)
  and 1 (bottom line).  The averaged double white dwarfs background is
  plotted as the solid line.  }
\label{fig:fh}
\end{figure*}

The total number of resolved systems with a signal above the sensitivity
limit (S/N = 1) of LISA for an assumed integration time of 1 yr is
12171. In Table~\ref{tab:resolved} we give the numbers of binaries of
different types that are resolved. The eccentric binaries can
contribute to more than one frequency [Eq.~(\ref{eq:h})], but the
amplitude of the signal in the high harmonics in general is rather
low. Just below the LISA sensitivity limit for $T$ = 1 yr there are
indeed 3 high harmonics of (ns, wd) binaries and one of a double
neutron star in our model.

The change of the frequency of a binary evolving under the influence
of GWR is given by \citep[e.g.][]{sch96}
\begin{equation}\label{eq:dotf}
\dot f = 5.8 \times 10^{-7} (\mathcal{M}/\msun)^{5/3} f^{11/3} \; {\rm Hz\,s^{-1}}.
\end{equation}  
For high-frequency systems this means that during a sufficiently long
integration time the change of the frequency (the ``chirp'') can be
detected and hence, $\mathcal{M}$ be determined. Then from
Eq.~(\ref{eq:h}) the distance to these sources can be found.  In
Fig.~\ref{fig:fMch} we plot the frequency versus chirp mass
distributions of the resolved systems for the binaries of different
types to show their properties separately. In the panel with the
double white dwarf systems we plot the so called chirp line, for $T =
1$~yr, as the solid line. Systems to the right of this line change
their frequency during the integration by one or more bins ($ \dot{f}
\, T \ge \Delta f = 1/T$).  The position at which the systems merge is
plotted as the dotted line in the left two panels.

The horizontal straight dotted line in Fig.~\ref{fig:fMch} (left
panel) marks the lower limit of $\mathcal{M}$ for systems with a total
mass larger than the Chandrasekhar mass which may be type Ia
Supernovae (SNe Ia) precursors. The resolved systems above this line
are plotted as the dots, since because of their relatively small
number (501 systems) the grey shades above this line are practically
invisible in the plot.

In the (ns, wd) panel we also plot the chirp and merger line, assuming
for the latter that the systems emit at the fundamental ($n = 2$)
frequency. In our model the mass of a neutron star is between $1.25$
and $1.55 \msun$, depending on the initial mass of the progenitor.
This results in a very narrow range in chirp masses for the double
neutron star systems.

The quantities measured by detectors like LISA are the frequency and
the strain amplitude [Eq.~(\ref{eq:h})]. In Fig.~\ref{fig:fh} we plot
the distributions of the expected resolved systems over $\log f$ and
$\log h$. In Fig.~\ref{fig:fh} we also show the sensitivity limits of
LISA for monochromatic sources, for signal to noise ratios of 5 and 1
and an integration time of 1 yr \citep[adapted from Fig.~5 and
Eq.~(53) of][]{lhh00}. The solid line gives the averaged noise
background as produced by double white dwarfs (see
Fig.~\ref{fig:GWR_bg}).

\subsection{Other detectable systems?}\label{detectable}

It may well be that the systems which produce strong signals because
of their proximity to the Sun or a large chirp mass can be detected
individually above the noise background (see Fig.~\ref{fig:GWR_bg}).
To investigate this possibility, we computed the number of systems of
the different types that are not resolved, but have a strain amplitude
well above the noise background, by selecting all systems that
  have a signal above the S/N = 5 sensitivity limit of LISA (see
  Fig.~\ref{fig:GWR_bg}). This adds a considerable number of
potentially detectable systems to the resolved binaries
(Table~\ref{tab:resolved}). It brings the total number of potentially
detectable binaries containing neutron stars to almost 200 for an
integration time of 1 yr.

\section{Discussion}\label{discussion}

\subsection{Halo and extra-galactic sources}

Above, the confusion limit and the number of resolved sources
were calculated for the Galactic disk binaries.  However, for example,
Galactic halo objects and extragalactic binaries may also contribute
to the GW signal in the LISA band.

The results of microlensing experiments can be considered as evidence
for the existence of massive compact halo objects (MACHO's).  The most
likely MACHO mass is between 0.15 and 0.9\,\msun, depending on the
halo model, and the total mass in MACHO's out to 50 kpc is
$9^{+4}_{-3} \times 10^{10}\,\msun$, independent of the halo model
\citep{aaa+00}.  The nature of MACHO's is still unknown
\citep[e.g.][]{ggh+98}, but two possibilities are relevant to this
study.

The first is that they are white dwarfs \citep{tsw+90}.  A discussion
is still going on whether the presence of a significant white dwarf
population is compatible with constraints derived from the chemical
composition of the halo and the cooling properties of white dwarfs
\citep[e.g.][]{cha99,ffg00,han00}. The fraction of binaries in this
hypothetical population is unknown.  However, because the star
formation in the halo happened long ago \citep[e.g.][]{al96}, most
short period double white dwarfs that could have formed will already
have merged. For example, for a 1\,Gyr long burst of star formation,
10\,Gyr ago, with an IMF similar to the IMF of the disk, all halo
close binary white dwarfs currently have orbital periods longer than
$\simeq 0.3$\,hr, so they cannot contribute to the GW signal at
frequencies higher than $\log f \simeq -2.75$.  The observed deficit
of halo white dwarf progenitors in distant galaxies \citep{al96}
suggests that the IMF in the halo is peaked at or above 2\,\msun,
limiting a hypothetical double white dwarf population to even lower
frequencies.  Hence, we do not expect a change in the confusion limit
due to halo white dwarfs, but they could contribute to the noise below
this limit.

Existing estimates of the contribution of halo double white dwarfs to
the GW noise \citep{hlr+00} are based on a simple rescaling to the
halo of the estimate of the GW signal from the disk by \citet{hbw90}.
Because of the different evolutionary histories of disk and halo this
is unrealistic.  Additionally, \citet{hlr+00} use the lowest existing
observational estimate of the local white dwarf space density,
probably overestimating the relative importance of a possible halo
white dwarf population.

Another possibility is that MACHO's are low-mass black holes.
Formation of low-mass ($\la 1\,\msun$) black holes is possible in
inflationary cosmological models \citep[e.g.][]{np85}. Further, as 
was shown by \citet{nst+97}, these black holes may form binaries.  An
estimate by \citet{his98} shows that under certain assumptions about the
separations of the components, the GW background formed by halo binary
black holes can be much stronger than the signal from the Galactic
binary white dwarfs. If this model is correct, the noise produced by
halo objects would bury virtually all resolved signals from Galactic
systems \citep[compare our Fig.\ref{fig:GWR_bg}, left panel and Fig.~2
of][]{his98}. This would result in the non-detection of any resolved
systems by LISA, and show up as an anisotropic noise.

A significant contribution from extragalactic binaries to the
background is expected only if the star formation keeps increasing at
$z \ga 3$ \citep{kp98}. A computation with a star formation rate which
is almost constant at $1.5 \la z \leq 5$ and roughly the same input
for stellar evolution as used in this study \citep{sfm+00}, showed
that only just above the point where the Galactic disk double white
dwarf background drops sharply (around $\log f = -2.75$) the
extragalactic background could exceed the Galactic one. However, the
signals of the resolved binaries at these frequencies are at least an
order of magnitude stronger than this background and will probably be
detectable \citep[see our Fig.~\ref{fig:fh} and Fig.~12 of][]{sfm+00}.

Finally it should be noted that, in addition to our estimate above, a
few tens of double neutron stars and neutron star -- white dwarf
binaries in globular clusters will probably be resolved \citep{bpr01}.

\subsection{Comparison with previous studies}

The Galactic GW background produced by double white dwarfs was studied
earlier by e.g. \citet{eis87,lpp87,hbw90,pp98,wh98}. The most widely
quoted study is the one by \citet{hbw90}, who calculated the
background based on the estimates of the number of systems by
\citet{web84}.  Because Webbink found a considerably higher birth rate
of close double white dwarfs than we find, Hills et al. found a higher
noise background.  The same holds for the study by \citet{eis87}, who
in addition use a different Galactic distribution.  When we estimate
the difference in the total number of systems in the Galaxy from the
different birth rates, we find that the renormalised background levels
differ by a factor $\la 3$.
 
\citet{wh98} use a current birth rate of 0.03~yr$^{-1}$, similar to
ours, but a constant star formation history and a larger age of the
Galactic disk, of 15 Gyr \citep{han98}. Their estimate of the
background is slightly higher that what we find, probably due to the
higher average chirp mass (0.42) of their (wd, wd) systems or their
assumed Galactic distribution, which is slightly different from our
Eq.~(\ref{eq:rho}).  Because they also calculated individual systems
(which they later average), they calculate the number of resolved
systems (although with a different criterium).  Above the resolution
limit found by Han \& Webbink ($\log f = -2.44$) we have 3615 resolved
systems in agreement with their number of 3600, despite differences in
the underlying white dwarf population.

The fact that most studies give an estimate of the confusion limit
within a factor $\sim2$ from $\sim 1.6 $\,mHz found by us is a
consequence of weak dependence of this limit on the parameters of the
models [see Eq. (19) of \citet{eis87}].

The background due to semi-detached white dwarfs was calculated by
\citet{hb00}. They conclude (as we do) that these stars are not
important for the overall background. This can be understood as a
consequence of their low strain amplitude due to their low chirp mass,
which makes them unimportant even in the frequency range where they
outnumber the detached systems.

\section{Conclusion}\label{conclusion}

We calculated the gravitational wave signal of Galactic disk binaries
containing two compact objects. We discuss three populations: (i)
double white dwarfs (including semi-detached systems) which produce a
confusion limited noise background at low frequencies ($\log f \la
-2.8$), (ii) resolved binaries and (iii) unresolved systems that have
such a strong signal that they may be detected above the noise.

The confusion limited background is dominated by detached double white
dwarfs, although in a small frequency range ($-3.4 \la \log f \la -3.0$)
the semi-detached systems form the majority of the systems in each bin.
The double white dwarf gravitational wave background, which in our
model consists of the sum of the signal of all 150 million systems in
the Galaxy shows large spikes caused by strong-signal
(i.e.  close) systems, which might be detectable above the noise.

Adding the binaries containing neutron stars and black holes (which
are much less numerous than white dwarf pairs), we find the
distribution of bins containing one, none and more than one system and
show that the ``confusion limit'' as a single value does not exist: at
$\log f = -2.84$ the first resolved bins are found, while up to $\log
f = -2.3$ bins containing more than one system are present.

We find 12171 resolved systems of which the vast majority are double
white dwarfs. There are only 8 double neutron stars and 38 neutron
star - white dwarf binaries resolved. Finally we calculate that there
are 6104 systems (5943 double white dwarfs, 124 neutron star -- white
dwarf systems, 31 double neutron stars and 6 systems containing a
black hole) which have a signal well above the double white dwarf
background and the LISA sensitivity level and might be detectable.

Out of 12124 resolved double white dwarfs, 501 have a combined mass
above the Chandrasekhar limit and periods short enough to merge in
10\,Gyr and are thus potential SN Ia progenitors. Such double white
dwarfs have not yet been found optically.  If a system would chirp,
LISA will measure the chirp mass and the distance to the system
allowing a good estimate of its total mass.  The number of chirping SN
Ia progenitors is $\sim94$ for an integration time of 1 yr.  But even
for these systems the actual time to coalescence is $\sim (4500 -
80000)$~yr.

\begin{acknowledgements}
  We thank Frank Verbunt for stimulating discussions.  LRY and SPZ
  acknowledge the warm hospitality of the Astronomical Institute
  ``Anton Pannekoek''. This work was supported by NWO Spinoza grant
  08-0 to E.~P.~J.~van den Heuvel, RFBR grant 99-02-16037, the Russian
  Ministry of Science Program ``Astronomy and Space Research'' and by
  NASA through Hubble Fellowship grant HF-01112.01-98A awarded (to
  SPZ) by the Space Telescope Science Institute, which is operated by
  the Association of Universities for Research in Astronomy, Inc., for
  NASA under contract NAS\,5-26555.
\end{acknowledgements}

\bibliography{journals,binaries}

\begin{thebibliography}{70}
\expandafter\ifx\csname natexlab\endcsname\relax\def\natexlab#1{#1}\fi

\bibitem[{{Adams} \& {Laughlin}(1996)}]{al96}
{Adams}, F.~C. \& {Laughlin}, G. 1996, \apj, 468, 586

\bibitem[{{Alcock} {et~al.}(2000){Alcock}, {Allsman}, {Alves}, {Axelrod},
  {Becker}, {Bennett}, {Cook}, {Dalal}, {Drake}, {Freeman}, {Geha}, {Griest},
  {Lehner}, {Marshall}, {Minniti}, {Nelson}, {Peterson}, {Popowski}, {Pratt},
  {Quinn}, {Stubbs}, {Sutherland}, {Tomaney}, {Vandehei}, \& {Welch}}]{aaa+00}
{Alcock}, C., {Allsman}, R.~A., {Alves}, D.~R., {et~al.} 2000, \apj, 542, 281

\bibitem[{Arzoumanian {et~al.}(1999)Arzoumanian, Cordes, \& Wasserman}]{acw99}
Arzoumanian, Z., Cordes, J.~M., \& Wasserman, I. 1999, \apj, 520, 696

\bibitem[{Barone {et~al.}(1988)Barone, Milano, Pinto, \& Recano}]{bmp+88}
Barone, F., Milano, L., Pinto, I., \& Recano, F. 1988, \aap, 199, 161

\bibitem[{Benacquista {et~al.}(2001)Benacquista, Portegies~Zwart, \&
  Rasio}]{bpr01}
Benacquista, M.~J., Portegies~Zwart, S., \& Rasio, F.~A. 2001, to appear in
  Class. Quantum Grav., gr-qc/0010020

\bibitem[{Brown {et~al.}(2001)Brown, Lee, Portegies~Zwart, \& Bethe}]{blp+01}
Brown, G.~E., Lee, C.-H., Portegies~Zwart, S.~F., \& Bethe, H. 2001, \apj, 547,
  345

\bibitem[{Cappellaro {et~al.}(1999)Cappellaro, Evans, \& Turatto}]{cet99}
Cappellaro, E., Evans, R., \& Turatto, M. 1999, A\&A, 351, 459

\bibitem[{{Chabrier}(1999)}]{cha99}
{Chabrier}, G. 1999, \apjl, 513, L103

\bibitem[{Charles(1998)}]{cha98}
Charles, P.~A. 1998, in Theory of Black Hole Accretion Disks, ed.
  M.~Abramowicz, G.~Bjornsson, \& J.~Pringle (Cambridge: CUP), 1

\bibitem[{Douglas \& Braginsky(1979)}]{db79}
Douglas, D.~H. \& Braginsky, V.~B. 1979, in General Relativity {\em An Einstein
  centenary survey}, ed. S.~W. Hawking \& W.~Israel (Cambridge: CUP), 90

\bibitem[{Eggleton {et~al.}(1989)Eggleton, Fitchett, \& Tout}]{eft89}
Eggleton, P.~P., Fitchett, M.~J., \& Tout, C.~A. 1989, ApJ, 347, 998

\bibitem[{Ergma \& van~den Heuvel(1998)}]{eh98a}
Ergma, E. \& van~den Heuvel, E. P.~J. 1998, A\&A, 331, L29

\bibitem[{Evans {et~al.}(1987)Evans, Iben, \& Smarr}]{eis87}
Evans, C.~R., Iben, Jr, I., \& Smarr, L. 1987, ApJ, 323, 129

\bibitem[{{Fields} {et~al.}(2000){Fields}, {Freese}, \& {Graff}}]{ffg00}
{Fields}, B.~D., {Freese}, K., \& {Graff}, D.~S. 2000, \apj, 534, 265

\bibitem[{Flanagan(1998)}]{fla98}
Flanagan, E. 1998, in Gravitation and Relativity: At the Turn of the
  Millennium, 177, gr-qc/9804024

\bibitem[{Fryer(1999)}]{fry99}
Fryer, C.~L. 1999, \apj, 522, 413

\bibitem[{Gates {et~al.}(1998)Gates, Gyuk, Holder, \& Turner}]{ggh+98}
Gates, E.~I., Gyuk, G., Holder, G.~P., \& Turner, M.~S. 1998, \apjl, 500, L145

\bibitem[{Han(1998)}]{han98}
Han, Z. 1998, MNRAS, 296, 1019

\bibitem[{Hansen(2000)}]{han00}
Hansen, B. M.~S. 2000, in Microlensing 2000: A new era of microlensing
  astrophysics, ed. J.~W. Menzies \& P.~D. Sackett, ASP Conf. Ser.

\bibitem[{Hartman(1997)}]{har97b}
Hartman, J.~W. 1997, A\&A, 322, 127

\bibitem[{Hils(1991)}]{hil91}
Hils, D. 1991, \apj, 381, 484

\bibitem[{Hils \& Bender(2000)}]{hb00}
Hils, D. \& Bender, P.~L. 2000, ApJ, 537, 334

\bibitem[{Hils {et~al.}(1990)Hils, Bender, \& Webbink}]{hbw90}
Hils, D., Bender, P.~L., \& Webbink, R.~F. 1990, ApJ, 360, 75

\bibitem[{Hiscock(1998)}]{his98}
Hiscock, W.~A. 1998, \apjl, 509, L101

\bibitem[{Hiscock {et~al.}(2000)Hiscock, Larson, Routzahn, \& Kulick}]{hlr+00}
Hiscock, W.~A., Larson, S.~L., Routzahn, J.~R., \& Kulick, B. 2000, \apjl, 540,
  L5

\bibitem[{Hulse \& Taylor(1975)}]{ht75}
Hulse, R.~A. \& Taylor, J.~H. 1975, \apjl, 195, L51

\bibitem[{Kalogera \& Lorimer(2000)}]{kl00}
Kalogera, V. \& Lorimer, D.~R. 2000, \apj, 530, 890

\bibitem[{Kalogera {et~al.}(2000)Kalogera, Narayan, Spergel, \&
  Taylor}]{kns+00}
Kalogera, V., Narayan, R., Spergel, D.~N., \& Taylor, J.~H. 2000, ApJ,
  submitted, astro-ph/0012038

\bibitem[{Kosenko \& Postnov(1998)}]{kp98}
Kosenko, D.~I. \& Postnov, K.~A. 1998, \aap, 336, 786

\bibitem[{Langer(1989)}]{lan89b}
Langer, N. 1989, A\&A, 220, 135

\bibitem[{Larson {et~al.}(2000)Larson, Hiscock, \& Hellings}]{lhh00}
Larson, S.~L., Hiscock, W.~A., \& Hellings, R.~W. 2000, Phys. Rev. D, 62,
  062001

\bibitem[{Lipunov {et~al.}(1987)Lipunov, Postnov, \& Prokhorov}]{lpp87}
Lipunov, V.~M., Postnov, K.~A., \& Prokhorov, M.~E. 1987, A\&A, 176, L1

\bibitem[{Lipunov {et~al.}(1997)Lipunov, Postnov, \& Prokhorov}]{lpp97}
---. 1997, Pisma Astronomicheskii Zhurnal, 23, 563

\bibitem[{{McNamara} {et~al.}(2000){McNamara}, {Ward}, \& {Hough}}]{mwh00}
{McNamara}, P.~W., {Ward}, H., \& {Hough}, J. 2000, Advances in Space Research,
  25, 1137

\bibitem[{Mironovskii(1965)}]{mir65}
Mironovskii, V.~N. 1965, SvA, 9, 752

\bibitem[{Nakamura {et~al.}(1997)Nakamura, Sasaki, Tanaka, \& Thorne}]{nst+97}
Nakamura, T., Sasaki, M., Tanaka, T., \& Thorne, K.~S. 1997, \apjl, 487, L139

\bibitem[{Naselskii \& Polnarev(1985)}]{np85}
Naselskii, P.~D. \& Polnarev, A.~G. 1985, SvA, 29, 487

\bibitem[{Nelemans {et~al.}(2000{\natexlab{a}})Nelemans, Portegies~Zwart, \&
  Verbunt}]{npv00}
Nelemans, G., Portegies~Zwart, S.~F., \& Verbunt, F. 2000{\natexlab{a}}, in
  Gravitational Waves and Experimental Gravity, ed. J.~Tr\^an Thanh~V\^an,
  J.~Dumarchez, S.~Reynaud, C.~Salomon, S.~Thorsett, \& J.~Y. Vinet, XXXIVth
  Rencontres de Moriond (Hanoi: World Publishers), 119

\bibitem[{Nelemans {et~al.}(2001{\natexlab{a}})Nelemans, Portegies~Zwart,
  Verbunt, \& Yungelson}]{npv+00}
Nelemans, G., Portegies~Zwart, S.~F., Verbunt, F., \& Yungelson, L.~R.
  2001{\natexlab{a}}, A\&A, 368, 939

\bibitem[{Nelemans {et~al.}(1999)Nelemans, Tauris, \& van~den Heuvel}]{nth99}
Nelemans, G., Tauris, T.~M., \& van~den Heuvel, E. P.~J. 1999, A\&A, 352, L87

\bibitem[{Nelemans {et~al.}(2000{\natexlab{b}})Nelemans, Verbunt, Yungelson, \&
  Portegies~Zwart}]{nvy+00}
Nelemans, G., Verbunt, F., Yungelson, L.~R., \& Portegies~Zwart, S.~F.
  2000{\natexlab{b}}, A\&A, 360, 1011

\bibitem[{Nelemans {et~al.}(2001{\natexlab{b}})Nelemans, Yungelson,
  Portegies~Zwart, \& Verbunt}]{nyp+00}
Nelemans, G., Yungelson, L.~R., Portegies~Zwart, S.~F., \& Verbunt, F.
  2001{\natexlab{b}}, A\&A, 365, 491

\bibitem[{Peters \& Matthews(1963)}]{pm63}
Peters, P.~C. \& Matthews, J. 1963, Phys.\ Rev., 131, 435

\bibitem[{Phinney(1991)}]{phi91}
Phinney, E.~S. 1991, \apjl, 380, L17

\bibitem[{Portegies~Zwart \& McMillan(2000)}]{pm00}
Portegies~Zwart, S.~F. \& McMillan, S. L.~W. 2000, \apjl, 528, L17

\bibitem[{Portegies~Zwart \& Verbunt(1996)}]{pv96}
Portegies~Zwart, S.~F. \& Verbunt, F. 1996, A\&A, 309, 179

\bibitem[{Portegies~Zwart {et~al.}(1997)Portegies~Zwart, Verbunt, \&
  Ergma}]{pve97}
Portegies~Zwart, S.~F., Verbunt, F., \& Ergma, E. 1997, \aap, 321, 207

\bibitem[{Portegies~Zwart \& Yungelson(1998)}]{py98}
Portegies~Zwart, S.~F. \& Yungelson, L.~R. 1998, A\&A, 332, 173

\bibitem[{Portegies~Zwart \& Yungelson(1999)}]{py99}
---. 1999, \mnras, 309, 26

\bibitem[{Postnov \& Prokhorov(1998)}]{pp98}
Postnov, K.~A. \& Prokhorov, M.~E. 1998, \apj, 494, 674

\bibitem[{Press \& Thorne(1972)}]{pt72}
Press, W. \& Thorne, K.~S. 1972, ARA\&A, 10, 335

\bibitem[{{Rana}(1991)}]{ran91}
{Rana}, N.~C. 1991, ARA\&A, 29, 129

\bibitem[{{Sackett}(1997)}]{sac97}
{Sackett}, P.~D. 1997, ApJ, 483, 103

\bibitem[{{Schaller} {et~al.}(1992){Schaller}, {Schaerer}, {Meynet}, \&
  {Maeder}}]{ssm+92}
{Schaller}, G., {Schaerer}, D., {Meynet}, G., \& {Maeder}, A. 1992, \aaps, 96,
  269

\bibitem[{Schneider {et~al.}(2000)Schneider, Ferrari, Matarrese, \&
  Portegies~Zwart}]{sfm+00}
Schneider, R., Ferrari, V., Matarrese, S., \& Portegies~Zwart, S.~F. 2000,
  MNRAS, submitted, astro-ph/0002055

\bibitem[{Schutz(1996)}]{sch96}
Schutz, B.~F. 1996, Class. Quantum Grav., 13, A219

\bibitem[{Tagoshi {et~al.}(2001)Tagoshi, Kanda, Tanaka, Tatsumi, Telada,
  {et~al.}}]{tkt+00}
Tagoshi, H., Kanda, N., Tanaka, T., {et~al.} 2001, \prd, 63, 231

\bibitem[{Tamanaha {et~al.}(1990)Tamanaha, Silk, Wood, \& Winget}]{tsw+90}
Tamanaha, C.~M., Silk, J., Wood, M.~A., \& Winget, D.~E. 1990, \apj, 358, 164

\bibitem[{Tauris \& Sennels(2000)}]{ts00}
Tauris, T.~M. \& Sennels, T. 2000, \aap, 355, 236

\bibitem[{Taylor \& Weisberg(1982)}]{tw82}
Taylor, J.~H. \& Weisberg, J.~M. 1982, \apj, 253, 908

\bibitem[{Tutukov \& Yungelson(1993{\natexlab{a}})}]{ty93c}
Tutukov, A.~V. \& Yungelson, L.~R. 1993{\natexlab{a}}, ARep, 37, 411

\bibitem[{Tutukov \& Yungelson(1993{\natexlab{b}})}]{ty93b}
---. 1993{\natexlab{b}}, \mnras, 260, 675

\bibitem[{van~den Heuvel \& Taam(1984)}]{ht84}
van~den Heuvel, E. P.~J. \& Taam, R.~E. 1984, \nat, 309, 235

\bibitem[{van~den Hoek \& de~Jong(1997)}]{hj97}
van~den Hoek, L.~B. \& de~Jong, T. 1997, A\&A, 318, 231

\bibitem[{Webbink(1984)}]{web84}
Webbink, R.~F. 1984, ApJ, 277, 355

\bibitem[{Webbink \& Han(1998)}]{wh98}
Webbink, R.~F. \& Han, Z. 1998, in Laser Interferometer Space Antenna, AIP
  Conf. Proc. No. 456 (New York: AIP), 61

\bibitem[{Weber(1969)}]{web69}
Weber, J. 1969, Phys. Rev. Lett., 22, 1302

\bibitem[{{Wellstein} \& {Langer}(1999)}]{wl99}
{Wellstein}, S. \& {Langer}, N. 1999, \aap, 350, 148

\bibitem[{White \& van Paradijs(1996)}]{wp96}
White, N.~E. \& van Paradijs, J. 1996, ApJ, 473, L25

\bibitem[{Woosley \& Weaver(1995)}]{ww95}
Woosley, S.~E. \& Weaver, T.~A. 1995, ApJS, 101, 181

\end{thebibliography}
\bibliographystyle{apj}

\appendix
\section{The formation of black holes}\label{bh_formation}

The simplified description of the evolution of massive stars used in
this paper is based on results of \citet{eft89} and \citet{ssm+92} and
can be summarised as follows \citep[see also][]{py98}
\begin{enumerate}
\item The radii of the massive stars are limited to 1000 $R_{\sun}$\
  \citep[after][]{ssm+92}.
\item The maximum amount of mass loss for hydrogen rich massive stars
  is given by 0.01 $M^2$, unless the whole envelope is lost and the
  star becomes a Wolf-Rayet star.  Thus stars above 85 \msun\ evolve
  off the main sequence immediately to become Wolf-Rayet stars with an
  initial mass of 43 \msun\ \citep[after][]{ssm+92}.
\item Wolf-Rayet stars lose mass according to the equation proposed by
  \citet{lan89b}: $\dot{M} = 5 \times 10^{-8} M^{2.5}$
  \msun~yr$^{-1}$.
\item In the supernova 50\% of the mass of the exploding object is
  ejected \citep[consistent with][]{nth99}.
\item Exploding helium stars more massive than 10\,\msun\ collapse to
  a black hole. Thus, the lower limit for black hole masses of $M = 5
  \msun$ is consistent with observational estimates
  \citep[e.g.][]{cha98}.
\item Black holes do not get a kick at birth in contrast to neutron
  stars \citep[e.g.][]{wp96}.
\end{enumerate}

\end{document}